\documentstyle[11pt]{article}
\begin{document}
%\vspace{-1cm}
%
\title{\bf Fractional Fokker-Planck Equation and Oscillatory Behavior of Cumulant Moments  }
\author{N. Suzuki$^1$, M. Biyajima$^2$ \\
   \vspace{1mm}\\
 $^1$ {\it Matsusho Gakuen Junior College, Matsumoto 390-1295, Japan}\\
 $^2$ {\it Department of Physics, Shinshu University, Matsumoto 390-8621
Japan} } 
\date{\today}
\maketitle 
\begin{abstract}
The Fokker-Planck equation is considered, which is connected to the birth and death process with immigration by the Poisson transform. The fractional derivative in time variable is introduced into the Fokker-Planck equation.  From its solution (the probability density function), the generating function (GF) for the corresponding probability distribution is derived.  We consider the case when the GF reduces to that of the negative binomial distribution (NBD), if the fractional derivative is replaced to the ordinary one. Formulas of the factorial moment and the $H_j$ moment are derived from the GF.  The $H_j$ moment derived from the GF of the NBD decreases monotonously as the rank j increases. However, the $H_j$ moment derived in our approach oscillates, which is contrasted with the case of the NBD.  Calculated $H_j$ moments are compared with those given from the data in $p\bar{p}$ collisions, $e^+e^-$ and $ep$ collisions.  A phenomenological meaning of introducing the fractional derivative in time variable is discussed.
\end{abstract}

\section{Introduction} 

The negative binomial distribution is often used for the analysis of observed multiplicity distributions in high energy hadron-hadron ($hh$) and $e^+e^-$ collisions. The cumulant moment derived from the generating function of the negative binomial distribution (NBD) does not show oscillatory behaviors as the rank of the cumulant moment increases.  On the other hand, cumulant moments obtained from observed multiplicity distributions in $hh$ and $e^+e^-$ collisions show oscillatory behaviors~\cite{drem93,drem94}.
Those behaviors can be explained if multiplicity distributions truncated at the highest observed multiplicities are used for the calculation of $H_j$ moments: In $hh$ collisions, calculated results from the NBD and those from the modified NBD both fit the data well~\cite{naka96,ugoc95}. In $e^+e^-$ collisions, calculated $H_j$ moments by the use of the modified NBD describe the oscillatory behavior of the data well.  However, those by the NBD oscillate much weaker than the data, and cannot explain the behavior of the data~\cite{suzu96}.

The NBD and the MNBD are derived from the branching equations; the former is from a birth and death process with immigration, and the latter is from a pure birth (or birth and death process). In those branching equations, it is assumed that particles are produced instantaneously, in other words, without memory.

In high energy particle-particle collision processes, it is considered that a  proper time is needed for a secondary produced particle to behave as an independent particle from the parent particle after the collision of the parent with a target particle~\cite{lpm53}.  In high energy hadron-nucleus or lepton-nucleus collisions, this effect should be observed as a supression of multiplicity compared with the case of instantaneous collision, because the incident particle can collide with another target particle in the same nucleus within the proper time after the first collision.  This proper time is called the formation zone~\cite{bial84}, which means some memory effect should be existent in the high energy particle production processes.

In the branching equations, particles are assumed to be produced successively.  If a memory effect is introduced into the branching process, it will be very interesting what results come out.

  The birth and death process with immigration is described by the following equation,
 \begin{eqnarray}
    \frac{\partial P(n,t)}{\partial t} 
        &=& \lambda_0 \left[ P(n-1,t) - P(n,t) \right]   \nonumber \\
        &&+ \lambda_2 \left[ (n-1)P(n-1,t) - nP(n,t) \right] \nonumber \\
        &&+ \lambda_1 \left[ (n+1)P(n+1,t) - nP(n,t) \right],  
              \label{eq.int1}
 \end{eqnarray}
where $P(n,t)$ denotes the probability distribution that n particles are existent at time t,  $\lambda_0$ denotes an immigration rate, $\lambda_1$ a death rate and $\lambda_2$ a birth rate.
If the initial condition is taken as
 \begin{eqnarray*}
       P(n,t=0)=0,
 \end{eqnarray*}
the solution of equation (\ref{eq.int1}) becomes the NBD.

The probability density function (KNO scaling function) $\psi (z,t)$ is connected to the probability distribution (multiplicity distribution) $P(n,t)$ by the Poisson transform,
 \begin{eqnarray}
   P(n,t) = \frac{\langle n_0 \rangle ^n}{n!}
             \int^\infty_0 z^n \exp[-\langle n_0 \rangle z] \psi(z,t) dz
      \label{eq.int2a}.
 \end{eqnarray}
The KNO scaling function $\psi (z,t)$ is obtained from the multiplicity distribution $P(n,t)$ by the inverse Poisson transform,
 \begin{eqnarray}
    \psi(z,t) = \frac{\langle n_0 \rangle}{2\pi\alpha}
             \exp[\langle n_0 \rangle z]
             \int^\infty_{-\infty} \sum_{n=0}^{\infty}
      \left(\frac{ix}{\alpha}\right)^n P(n,t) 
          \exp[-ix\frac{\langle n_0 \rangle}{\alpha}z] dx
      \label{eq.int2b}.
 \end{eqnarray}
Applying the inverse Poisson transform (\ref{eq.int2b}) to Eq.(\ref{eq.int1}), we obtain the Fokker-Planck equation,
 \begin{eqnarray}
  \frac{\partial \psi (z,t)}{\partial t} = -\frac{\partial}{\partial z}
    \left[ a(z) - \frac{1}{2}\frac{\partial}{\partial z}b(z) \right]\psi (z,t),                                            \label{eq.int3} 
 \end{eqnarray}
where
 \begin{eqnarray}
      a(z) &=& \beta - \gamma z,  \quad b(z)= \sigma^2 z,   \nonumber \\
      \beta &=& \frac{\lambda_0}{\langle n_0 \rangle},  \quad
    \gamma = \lambda_1 - \lambda_2,   \quad
    \sigma^2 = \frac{2\lambda_2}{\langle n_0 \rangle}.      \label{eq.int4}        
 \end{eqnarray}
If time derivative in Eq.(\ref{eq.int3}) is replaced to the fractional one, we have reached to the fractional Fokker-Planck equation in time variable as a model for high energy particle production processes, in which a memory effect is taken into account. 

The fractional calculus has been investigated for hundreds of years~\cite{oldh74,podl99}.  Recently, the fractional Fokker-Planck equation in time variable was derived from the continuous time random walk~\cite{bark00a}.  It is applied to the analysis of anomalous diffusion phenomena~\cite{bark00b}.  The fractional derivative in space variable is introduced into the Fokker-Planck equation to describe the L\'{e}vy process~\cite{west97}. 

We would like to take the fractional Fokker-Planck equation in time variable corresponding to the branching equation (\ref{eq.int1}) as a model for particle production process, and to investigate it's solution, which reduces to the gamma distribution when the fractional derivative is replaced to the ordinary one. 
We also examine the effect of fractional derivative or introducing the memory effect on the behavior of cumulant moments.

\section{The Fractional Fokker-Planck Equation}

In the following we consider the fractional Fokker-Planck equation (FFPE), 
 \begin{eqnarray}
  \frac{\partial \psi (z,t)}{\partial t}
                &=& {}_0{\rm D}^{1-\alpha}_tL_{FP} \psi (z,t), 
                 \quad 0<\alpha<1, \nonumber \\
  L_{FP} &=& -\frac{\partial}{\partial z}
    \left[ a(z) - \frac{1}{2}\frac{\partial}{\partial z}b(z) \right],
                          \label{eq.frac1} 
 \end{eqnarray}
with the initial condition,
 \begin{eqnarray}
   \psi (z,t=0) = \delta(z-z_0) \label{eq.frac2}.
 \end{eqnarray}
In Eq.({\ref{eq.frac1}), ${}_0{\bf D}_t^\delta$ denotes the Riemann-Liouville fractional derivative~\cite{oldh74,podl99} defined by
 \begin{eqnarray}
   {}_0{\bf D}_t^\delta f(t) = \frac{1}{\Gamma(n-\delta)} \frac{d^n}{dt^n}
             \int^t_0 (t-\tau)^{n-\delta-1} f(\tau) d\tau, 
   \quad n-1 \le \delta <n,  \label{eq.frac3}
 \end{eqnarray}
where $n$ is a positive integer.  If $\alpha=1$, Eq.(\ref{eq.frac1}) reduces to Eq.(\ref{eq.int3}).

According to the method proposed by Barkai and Silbey~\cite{bark00b}, 
we assume that
 \begin{eqnarray}
       \psi(z,t) = \int_0^{\infty}  R_s(t) G_s(z) ds,   \label{eq.frac4}
 \end{eqnarray}
and that function $G_s(z)$ satisfies the following equations,
 \begin{eqnarray}
    L_{FP} G_s(z) = \frac{\partial}{\partial s} G_s(z), \nonumber \\
    G_0(z)=\delta(z-z_0).  \label{eq.frac5}
 \end{eqnarray}
Then function $G_s(z)$ is given as,
 \begin{eqnarray}
    G_s(z)= \frac{1}{kp} 
          \exp[-\frac{z+z_0(1-p)}{kp}]
          \left( \frac{z}{z_0(1-p)} \right)^{(\lambda -1)/2}
      I_{\lambda-1} ( \frac{2 \sqrt{zz_0(1-p)} }{kp} ),
                  \label{eq.frac6}
 \end{eqnarray}
where
 \begin{eqnarray}
     k= \frac{\sigma^2}{2\gamma},  \quad
    \lambda = \frac{2 \beta}{\sigma^2}, \quad
     p = 1 - e^{-\gamma s}.                    \label{eq.frac7}
 \end{eqnarray}
In the following, we assume that $\gamma = \lambda_1 - \lambda_2 > 0$.
Equation (\ref{eq.frac6}) can be expanded as
 \begin{eqnarray}
    G_s(z)= \frac{1}{k} ( \frac{z}{k} )^{\lambda -1} 
           \exp[-\frac{z}{k}]
           \sum_{m=0}^{\infty} \frac{m!}{\Gamma(m+\lambda)}
           L_m^{(\lambda-1)}(\frac{z}{k})  L_m^{(\lambda-1)}(\frac{z_0}{k}) 
           \exp[-m\gamma s],  \label{eq.frac8}
 \end{eqnarray}
where $L_m^{(\lambda-1)}(z)$ denotes the Laguerre polynomial.

Applying the Laplace transform to Eq.(\ref{eq.frac1}), we find
 \begin{eqnarray}
     \int_0^\infty \left[ u \tilde R_s(u)
        +  u^{1-\alpha}\frac{\partial \tilde R_s(u)}{\partial s} 
            \right] G_s(z) ds
       = \left[ 1 - u^{1-\alpha}\tilde R_0(u) \right]\delta(z-z_0), 
         \label{eq.frac9}
 \end{eqnarray}
where $\tilde R_s(u)$ is the Laplace transform of $R_s(t)$,
 \begin{eqnarray}
   \tilde R_s(u) = \int_0^{\infty} R_s(t) e^{-ut} dt. 
                 \label{eq.frac10}
 \end{eqnarray}

Furthermore, we assume that each side of Eq.(\ref{eq.frac9}) is equal to zero;
 \begin{eqnarray}
    u^{1-\alpha}\tilde R_0(u) = 1,  \quad 
    -u^{1-\alpha} \frac{\partial \tilde R_s(u)}{\partial u} = u\tilde R_s(u).
               \label{eq.frac11}
 \end{eqnarray}
The solution of Eq.(\ref{eq.frac11}) is given by
 \begin{eqnarray}
    \tilde R_s(u) = u^{\alpha-1} \exp[-su^\alpha].  \label{eq.frac12}
 \end{eqnarray}
Then $R_s(t)$, the inverse Laplace transform of $\tilde R_s(u)$, is written as
 \begin{eqnarray}
   R_s(t) &=& \frac{1}{2\pi i} \int_{c_0-i\infty}^{c_0+i\infty}
                  \tilde R_s(u) e^{ut} du   \nonumber \\
          &=& \frac{t^{-\alpha}}{2\pi i}
              \int_{c_0-i\infty}^{c_0+i\infty}\sigma^{\alpha-1}
             \exp[\sigma - \frac{s}{t^\alpha} \sigma^\alpha] d \sigma,
                \quad (c_0>0).
                \label{eq.frac13}
 \end{eqnarray}

Therefore, the solution for the FFPE (\ref{eq.frac1}) is given by
 \begin{eqnarray}
    \psi(z,t) &=& \int_0^{\infty} R_s(t) G_s(z) ds    \nonumber \\
         &=& \frac{z^{\lambda -1}}{k^{\lambda}}  
           \exp[-\frac{z}{k}]
           \sum_{m=0}^{\infty} \frac{m!}{\Gamma(m+\lambda)}
           L_m^{(\lambda-1)}(\frac{z}{k})  L_m^{(\lambda-1)}(\frac{z_0}{k}) 
           E_{\alpha}(-m\gamma t^\alpha),
           \label{eq.frac14}
 \end{eqnarray}
where $E_\alpha(-t)$ for $t>0$ denotes the Mittag-Leffler function~\cite{gore96},
 \begin{eqnarray}
    E_\alpha(-t) = \frac{\sin(\alpha\pi)}{\alpha\pi} \int_0^\infty
           \exp[-(xt)^{1/\alpha}] \frac{1}{x^2+2x\cos(\alpha\pi)+1}dx.
                           \label{eq.frac15}
 \end{eqnarray}
It is written in the infinite series as
 \begin{eqnarray}
    E_\alpha(z) = \sum_{n=0}^{\infty} \frac{z^n}{\Gamma(\alpha n+1)}.
                          \label{eq.frac16}
 \end{eqnarray}

In the limit of $z_0 \rightarrow +0$,  Eq.(\ref{eq.frac14}) reduces to
 \begin{eqnarray}
    \psi(z,t) = \frac{1}{\Gamma(\lambda)}\frac{z^{\lambda -1}}{k^{\lambda}}
          \exp[-\frac{z}{k}] \sum_{m=0}^{\infty} 
      L_m^{(\lambda-1)}(\frac{z}{k}) E_{\alpha}(-m\gamma t^\alpha).
                            \label{eq.frac17}
 \end{eqnarray}
If $\alpha=1$, Eq.(\ref{eq.frac17}) coincides with the gamma distribution, the KNO scaling function of the NBD\footnote{
We have considered the FFPE for $\gamma = \lambda_1 - \lambda_2 > 0$, where the birth rate  $\lambda_2$ is less than the immigration rate $\lambda_1$.
In the case for $\gamma<0$, the solution of the FFPE (\ref{eq.frac1}) in the limit of $z_0 \rightarrow +0$,  is given by 
 \begin{eqnarray*}
   \psi(z,t) = \frac{1}{\Gamma(\lambda)}\frac{z^{\lambda -1}}{|k|^{\lambda}}
           \sum_{m=0}^{\infty} L_m^{(\lambda-1)}(\frac{z}{|k|})
           E_{\alpha}(-(m+\lambda)|\gamma| t^\alpha).
 \end{eqnarray*}
If $\alpha=1$, the above equation coincides with the gamma distribution.  However, we cannot calculate the factorial moment from it, because the exponential damping factor in z variable is not contained in the equation, contrary to Eq.(\ref{eq.frac17}).

}. 

\section{Factorial Moment and Cumulant Moment}

The generating function (GF) for the multiplicity distribution $P(n,t)$ is defined as,
 \begin{eqnarray}
    \Pi(u) = \sum_{n=0}^{\infty} P(n,t) u^n.  \label{eq.fact1}
 \end{eqnarray}
The multiplicity distribution and the $j$th rank factorial moment are given from Eq.(\ref{eq.fact1}) respectively as
 \begin{eqnarray}
    P(n,t) &=& \frac{1}{n!}\frac{\partial^n 
              \Pi(u)}{\partial u^n}\Big|_{u=0}, \label{eq.fact1b} \nonumber \\
       f_j &=& \langle n(n-1)\cdots (n-j+1)\rangle   \nonumber \\ 
           &=& \frac{\partial^j \Pi(u)}{\partial u^j}\Big|_{u=1}
          = \sum_{n=j}^{\infty}n(n-1)\cdots(n-j+1)P(n).   \label{eq.fact1c}
 \end{eqnarray}
From Eqs.(\ref{eq.int2a}) and (\ref{eq.fact1}), the GF is written by the use of the Laplace transform of the KNO scaling function $\psi(z,t)$ as
 \begin{eqnarray}
    \Pi( 1-{u}/{\langle n_0 \rangle} )
        = \int_0^\infty \psi(z,t) e^{-uz}dz.
                         \label{eq.fact2}
 \end{eqnarray}
Then the GF corresponding to Eq.(\ref{eq.frac17}) is given as,
 \begin{eqnarray}
    \Pi(u) = \sum_{m=0}^{\infty} 
       \frac{\Gamma(m+\lambda)}{m!\Gamma(\lambda)}
       \frac{ \left( -k\langle n_0 \rangle (u-1) \right)^m }
         {\left(1 -k\langle n_0 \rangle (u-1) \right)^{m+\lambda}}
            E_{\alpha}(-m\gamma t^\alpha).
                       \label{eq.fact3}
 \end{eqnarray}
The multiplicity distribution and the factorial moment are given from Eqs.(\ref{eq.fact1c}) and (\ref{eq.fact3}) respectively as,
 \begin{eqnarray}
    P(n,t) &=& \sum_{m=0}^{\infty} \sum_{l=0}^{{\rm min}(m,n)}
              \frac{(-1)^l\Gamma(m+n+\lambda-l)}
            {\Gamma(\lambda)(m-l)! (n-l)! l!}
                                        \nonumber \\
           &&\times \frac{ \left( k\langle n_0 \rangle \right)^{m+n-l} }
           { \left(1+k\langle n_0 \rangle \right)^{m+n+\lambda-l} }
            E_{\alpha}(-m\gamma t^\alpha),
                      \label{eq.fact4}   \\
    f_j &=& (k\langle n_0 \rangle)^j \frac{\Gamma(\lambda+j)}{\Gamma(\lambda)}
     \sum_{m=0}^{j} (-1)^m {}_jC_m  E_{\alpha}(-m\gamma t^\alpha).
                     \label{eq.fact5}
 \end{eqnarray}
The $j$th rank normalized factorial moment is given by
 \begin{eqnarray}
    F_j =\frac{f_j}{\langle n \rangle^j}
        = \frac{\Gamma(\lambda+j)}{\Gamma(\lambda)\lambda^j}
    \frac{\sum_{m=0}^{j}(-1)^m{}_jC_m 
          E_{\alpha}(-m\gamma t^\alpha)}
       {[1-E_\alpha(-\gamma t^\alpha)]^j}.   \label{eq.fact6}
 \end{eqnarray}
The $k$th rank cumulant moment is defined by the following equation,
 \begin{eqnarray}
    \kappa_j = \frac{\partial^j \ln\Pi(u)}{\partial u^j}\Big|_{u=1}.
                  \label{eq.fact7}
 \end{eqnarray}
From Eqs.(\ref{eq.fact5}), (\ref{eq.fact6}) and (\ref{eq.fact7}), we obtain the recurrence equation for the $H_j$ moment,
 \begin{eqnarray}
      H_1 = 1,  \quad
      H_j &=& 1 - \sum_{m=1}^{j-1}{}_{j-1}C_{m-1}\frac{F_{j-m}F_m}{F_j}H_m, 
                                        \label{eq.fact8}
 \end{eqnarray}
where    \[   H_j = {\kappa_j}/{f_j}.  \]

If $\alpha=1$, Eq.(\ref{eq.fact3}) reduces to the generating function for the NBD with mean multiplicity $\langle n_{\rm b} \rangle=k\lambda\langle n_0 \rangle(1-e^{-\gamma t})$,
\begin{eqnarray}
       \Pi_{\rm NB}(u) = \left[ 
         1- \frac{\langle n_{\rm b} \rangle}{\lambda}(u-1)
                    \right]^{-\lambda}.  \label{eq.fact9}
\end{eqnarray}
From Eq.(\ref{eq.fact9}), the NBD is given as
 \begin{eqnarray}
   P(n,t) = \frac{\Gamma(\lambda + n)}{\Gamma(\lambda) \Gamma(n+1)}
   \frac{(\langle n_{\rm b} \rangle/\lambda)^n }
        { (1 + \langle n_{\rm b} \rangle/\lambda)^{n+\lambda} }.
                    \label{eq.frac9b}
 \end{eqnarray}
The normalized factorial moment and the $H_j$ moment for the NBD are given respectively as,
\begin{eqnarray}
       F_{\rm NB},_j = \frac{\Gamma(\lambda+j)}
                      {\Gamma(\lambda)\lambda^j}, \quad
       H_{\rm NB},_j = \frac{\Gamma(\lambda+1)(j-1)!}{\Gamma(\lambda+j)}.
                   \label{eq.fact10}
\end{eqnarray}
As can be seen from Eqs.(\ref{eq.fact6}) and (\ref{eq.fact10}), difference between the normalized factorial moment derived from the FFPE ($0<\alpha<1$)  and that of the NBD ($\alpha=1$) is given by Mittag-Leffler functions.
We can see from Eq.(\ref{eq.frac15}) that
 \begin{eqnarray*}
    \lim_{t \rightarrow + \infty}  E_\alpha(-\gamma t^\alpha) = 0.
 \end{eqnarray*}
Then, equation (\ref{eq.fact6}) coincide with Eq.(\ref{eq.fact10}) in the limit of $t \rightarrow + \infty$.

\section{Calculated Results}

At first, calculated results of the Mittag-Leffler function $E_\alpha(-t)$ is shown in Fig.1. It is a decreasing function of variable $t$,  and as $\alpha$ increases from 0 to 1, it decreases more faster as a function of variable t.

In the following calculations, observed values of $\langle n \rangle$ and $C^2 (=\langle n^2 \rangle/\langle n \rangle^2)$ for the charged particles are used. Then, if $\alpha$ and $\gamma t^\alpha$ are  given, $\lambda$ in Eq.(\ref{eq.fact6}) is determined by the following equation, 
 \begin{eqnarray*}
    \frac{1}{\lambda}
        = \left(C^2 - \frac{1}{\langle n \rangle} \right)
          \frac{[1-E_\alpha(-\gamma t^\alpha)]^2}
               {1-2E_\alpha(-\gamma t^\alpha) + E_{\alpha}(-2\gamma t^\alpha)}
          -1.
                         \label{eq.calc1}
 \end{eqnarray*}

In order to see the effect of the fractional derivative, i.e. $0<\alpha<1$, to the oscillatory behavior of $H_j$ moments, calculated $H_j$ moments are shown in Fig.2a with $\alpha=$0.25, and $\gamma t^\alpha=$1.5, 2.0 and 2.5, in Fig.2b with $\alpha=$0.50, and $\gamma t^\alpha=$1.0, 1.5 and 2.0, and in Fig.2c with $\alpha=$0.75, and $\gamma t^\alpha=$0.5, 1.0 and 1.5. 
In our calculation, $\langle n \rangle=29.2$ and $C^2=1.274$, observed values in $p\bar{p}$ collisions at $\sqrt{s}=546$ GeV, are used~\cite{ua587}.
 If $\alpha$ is fixed, oscillation of the $H_j$ moment as a function of rank $j$ becomes weaker as the value of parameter $\gamma t^\alpha$ increases.  If $\gamma t^\alpha$ is fixed, oscillation of $H_j$ moments become much weaker as $ \alpha $ increases from 0 to 1.

 In Fig.3, our calculation with $\alpha=0.5$ and $\gamma t^\alpha=2.23$ is compared with the $H_j$ moment obtained from the data in $p\bar{p}$ collisions at $\sqrt{s}=546$GeV~\cite{ua587}.  Parameter $\gamma t^\alpha$ is adjusted with a step of 0.01 so that the first relative minimum of the calculated $H_j$ moment should be located near the rank of that obtained from the data as much as possible.
The calculated first relative minimum value is $H_7=-3.32\times 10^{-5}$, and the absolute value of it is much smaller than that obtained from the data.  However, we can see from Figs.2b and 3, the calculated $H_j$ moment with $\alpha=0.5$ and $\gamma t^\alpha=1.0$ oscillates as strong as that from the data.  

 In Fig.4, the calculated $H_j$ moment with $\alpha=0.5$ and $\gamma t^\alpha=90.0$ is compared with that obtained from the data in $e^+e^-$ collisions at $\sqrt{s}=91$ GeV~\cite{sld96}. As can be seen from the figure, the calculated value of the first relative minimum is almost the same with the data, and the strength of the oscillation of calculated $H_j$ moment is comparable with the data.

The $H_j$ moment in $ep$ collisions in the interval $185<W<220$ GeV~\cite{h196} is also analyzed, and the results are shown in Fig5. Calculated $H_j$ moment with $\alpha=$0.5 and $\gamma t^\alpha=$11.2 ($\lambda=$20.00) well reproduces the first relative minimum of the data. For comparison, the $H_j$ moment calculated with the NBD (\ref{eq.frac9b}) truncated at the highest observed charged multiplicity is also shown.  The parameters of the truncated NBD are determined by the minimum chi-squared fit with the observed charged multiplicity distribution. The first relative minimum of the calculated $H_j$ moment with the truncated NBD is $H_5=-3.42\times 10^{-3}$ which is different from the first relative minimum $H_3=-4.07\times 10^{-3}$ obtained from the experimental data. However the strength of the oscillation of $H_j$ moments calculated with the truncated NBD is comparable with the data.  
Estimation of $H_j$ moment is obtained by the use of Eq.(\ref{eq.fact8}) from the $F_j$ moment both in the theoretical calculation and in the calculation from the experimental data.  In order to see the relation between the behavior of the $H_j$ moment and that of the normalized factorial moment $F_j$ as a function of rank $j$, we also analyze the $F_j$ moment in $ep$ collisions in the interval $185<W<220$ GeV~\cite{h196}. The parameters are the same with those in Fig.5.  The calculated $F_j$ moment with Eq.(\ref{eq.frac10}) is compared with the data in Fig.6a. 
Our calculated result in the normalized factorial moment well reproduce the experimental data up to the fourth rank. The difference between them is less than one percent.  However we cannot reproduce the fourth rank $H_j$ moment of the data from our calculation.  This result indicates that the oscillation of the $H_j$ moment is very sensitive to the value of $F_j$ moments.

In order to see the effect of truncation for the normalized factorial moment $F_j$, those calculated with Eq.(\ref{eq.fact10}) (without truncation) and with truncated NBD are shown in Fig.6b.  The former is calculated with $\langle n \rangle=8.80$ and $C^2=1.190$. The parameters of the truncated NBD are the same with those in Fig.5.   The $F_j$ moment with truncation is much more suppressed than that without truncation at higher rank $j$.   

 From Figs.6a and 6b, we can see that the $F_j$ moment calculated from the FFPE is smaller than that with Eq.(\ref{eq.fact10}) (without truncation). Therefore, introduction of the fractional derivative in time variable suppresses the value of $F_j$ moment compared with that with Eq.(\ref{eq.fact10}) as the rank $j$ increases, and gives rise to similar effect as in truncation of multiplicity distribution.

The $H_j$ moments and normalized factorial moments in $ep$ collisions in the interval $185<W<220$ GeV~\cite{h196} are also calculated with sets of parameters,  $\alpha=$0.25 and $\gamma t^\alpha=$16.0 ($\lambda=$19.94), or $\alpha=$0.75 and $\gamma t^\alpha=$5.70 ($\lambda=$20.41).  The resullts become almost the same with those shown in Figs.5 and 6.

\section{Concluding Remarks}

The fractional Fokker-Planck equation (FFPE) corresponding to the birth and death process with immigration is taken as a model for particle production processes with a memory effect.  It is solved according to the procedure proposed by Barkai and Silbey\cite{bark00b}.
From the solution of the FFPE, we obtain the generating function for the multiplicity distribution, where parameter $\alpha$ connected with the fractional time derivative is contained. If $\alpha$ is put to 1, the distribution becomes the NBD.  

The normalized factorial moment $F_j$ calculated with Eq.(\ref{eq.frac6}) 
becomes much smaller than that with Eq.(\ref{eq.frac10}) obtained from the GF for the NBD as the rank $j$ increases, where Eq.(\ref{eq.frac6}) coincides with Eq.(\ref{eq.frac10}) if $\alpha=1$.  This fact means that the high multiplicity component in the multiplicity distribution is surppressed if $\alpha<1$,  compared with the NBD ($\alpha=1$) with the same $\langle n \rangle$ and $C^2$. 

When $\alpha$ is less than 1, the oscillation of $H_j$ moment appears, and as $\alpha$ decreases from 1 to 0, the oscillation becomes much stronger. 
This is caused by the fact that the fractional derivative ($0<\alpha<1$) is introduced into the time derivative in Eq.(\ref{eq.frac1}). It can be said that introducing the fractional derivative gives similar effect on the normalized factorial moment and the $H_j$ moment as in the truncation of multiplicity distributions.

\newpage
{\bf Figure captions}
\begin{itemize}
\item[{\bf Fig. 1}] The Mittag-Leffler function calculated from 
   Eq.(\ref{eq.frac15}) with $\alpha=$0.25, 0.50, 0.75 and 1.00.
\item[{\bf Fig. 2}] Calculated results of $H_j$ moments as a function of 
     rank $j$; a) with $\alpha$=0.25, and $\gamma t^\alpha=$1.5
               ($\lambda=$48.44), 2.5 (16.93) and 3.5 (11.61).
               b) with $\alpha$=0.50, and $\gamma t^\alpha=$1.0 
                  ($\lambda=$66.82), 1.5 (15.41) and 2.0 (10.24). 
               c) with $\alpha$=0.75, and $\gamma t^\alpha=$0.5 
                  ($\lambda=$20.41), 1.0 (10.20) and 1.5 (7.65).
     In each calculation, $\langle n \rangle=29.2$ and $C^2=1.274$, 
    observed values of charged particles in $p\bar{p}$ collisions at 
    $\sqrt{s}=$546 GeV/c~\cite{ua587} are used 
\item[{\bf Fig. 3}] Calculated result with  $\alpha=0.5$ and 
   $\gamma t^\alpha = 2.23$  ($\lambda = 9.16$) is compared 
    with the data of charged particles in $p\bar{p}$ collisions at 
    $\sqrt{s}=$546 GeV/c~\cite{ua587}, where $\langle n \rangle=29.2$ 
    and $C^2=1.274$ are used. 
     Parameter $\alpha$ is fixed at 0.5, and $\gamma t^\alpha$ is adjusted 
     so that the first relative minimum of the calculated $H_j$ moment is 
     located at $j \ge 5$ ($j=7$).
\item[{\bf Fig. 4}] Calculated $H_j$ moments with $\alpha$=0.50 and $\gamma t^\alpha=$90.0 ($\lambda = 25.36$), as a function of rank $j$  are 
     compared with those in $e^+e^-$ collisions at $\sqrt{s}=91$ 
     GeV/c~\cite{sld96}. $\langle n \rangle=20.70$ and $C^2=11.091$ are used in our calculation.
\item[{\bf Fig. 5}] Calculated $H_j$ moments as a function of rank $j$  are compared with the data in $ep$ collisions in the interval $185<W<220$ GeV~\cite{h196}.  White circles denote the calculated result with $\alpha$=0.50 and $\gamma t^\alpha=$11.2, where $\langle n \rangle=8.80$ and $C^2=1.190$ are used. White triangles show the $H_j$ moment calculated with the NBD which is truncated at $n=21$, the highest observed charged multiplicity, where $\langle n_{\rm b} \rangle=8.825$ and $\lambda=13.21$ are used.  
\item[{\bf Fig. 6}] Calculated normalized factorial moments $F_j$ as a function of rank $j$ are compared with the data in $ep$ collisions in the interval $185<W<220$ GeV~\cite{h196}.
  a) White circles denote the calculated result with $\alpha$=0.50 and $\gamma t^\alpha=$11.2 ($\lambda=20.00$), and black circles are the data. 
     
b)  White circles denote the result calculated with the NBD which is truncated at $n=21$, where $\langle n_{\rm b} \rangle=8.825$ and $\lambda=13.21$ are used.  Crosses show calculated $H_j$ moment from the GF, Eq.(\ref{eq.frac10}), where 
  $\lambda=13.10$ is used.
\end{itemize}
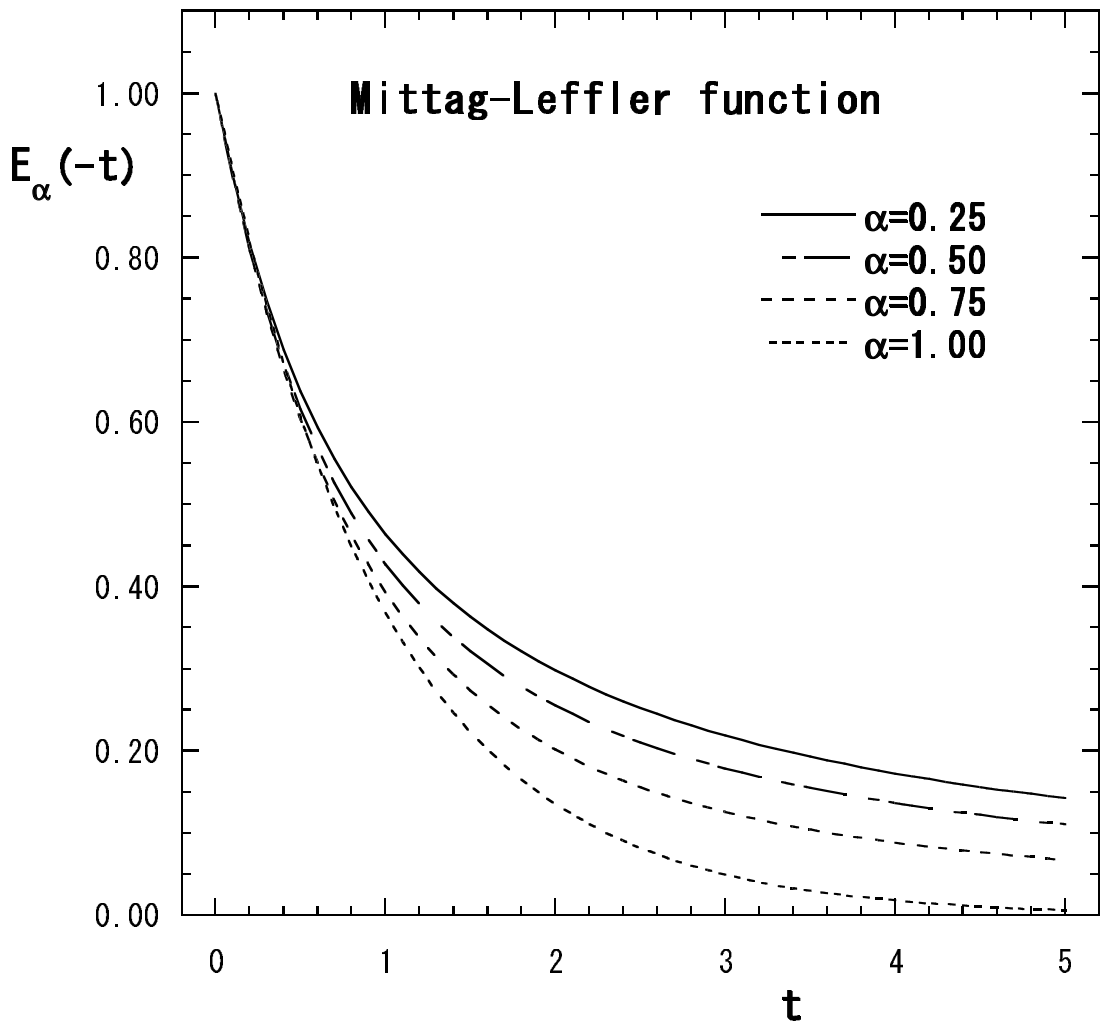
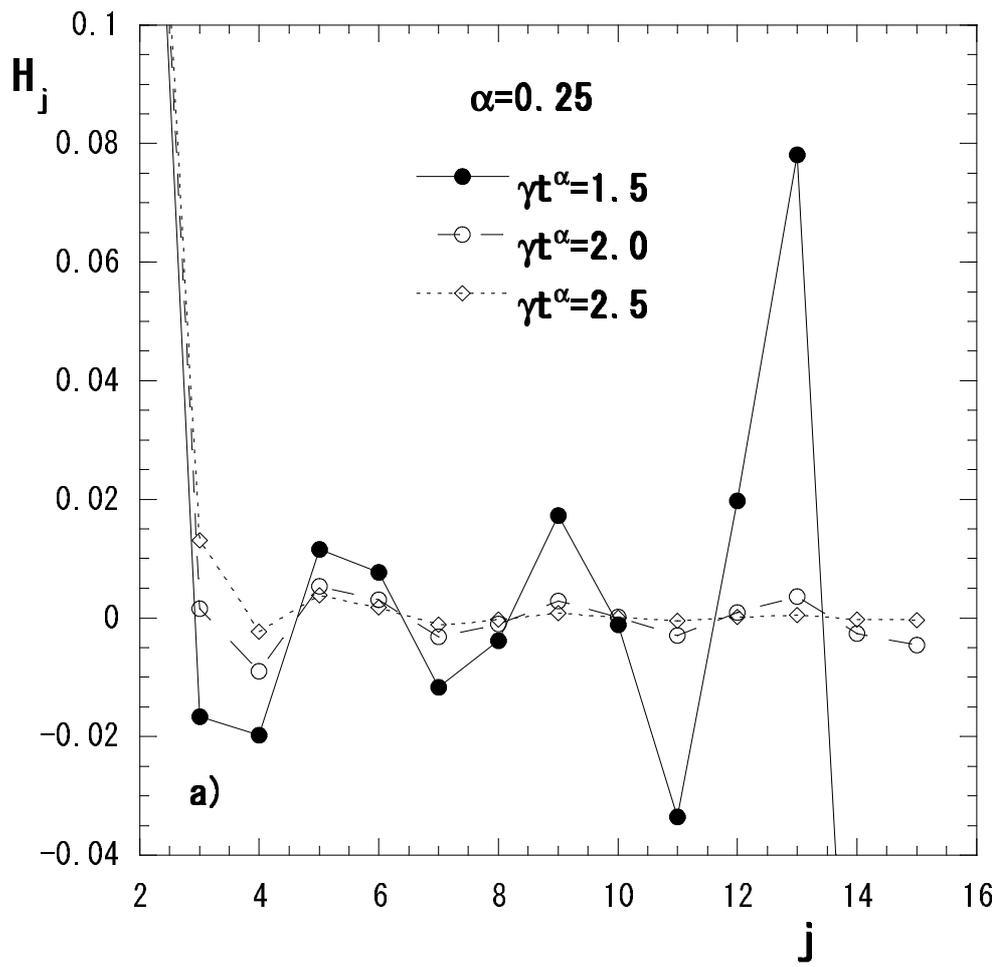
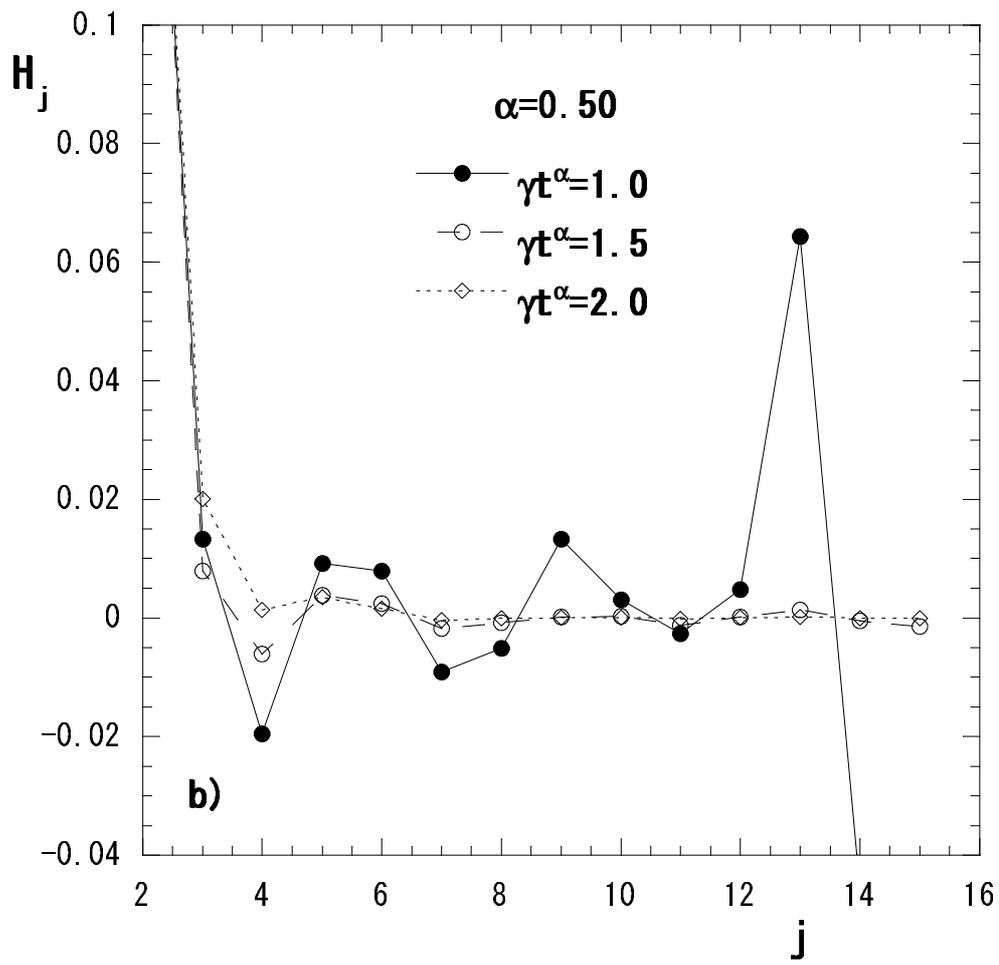
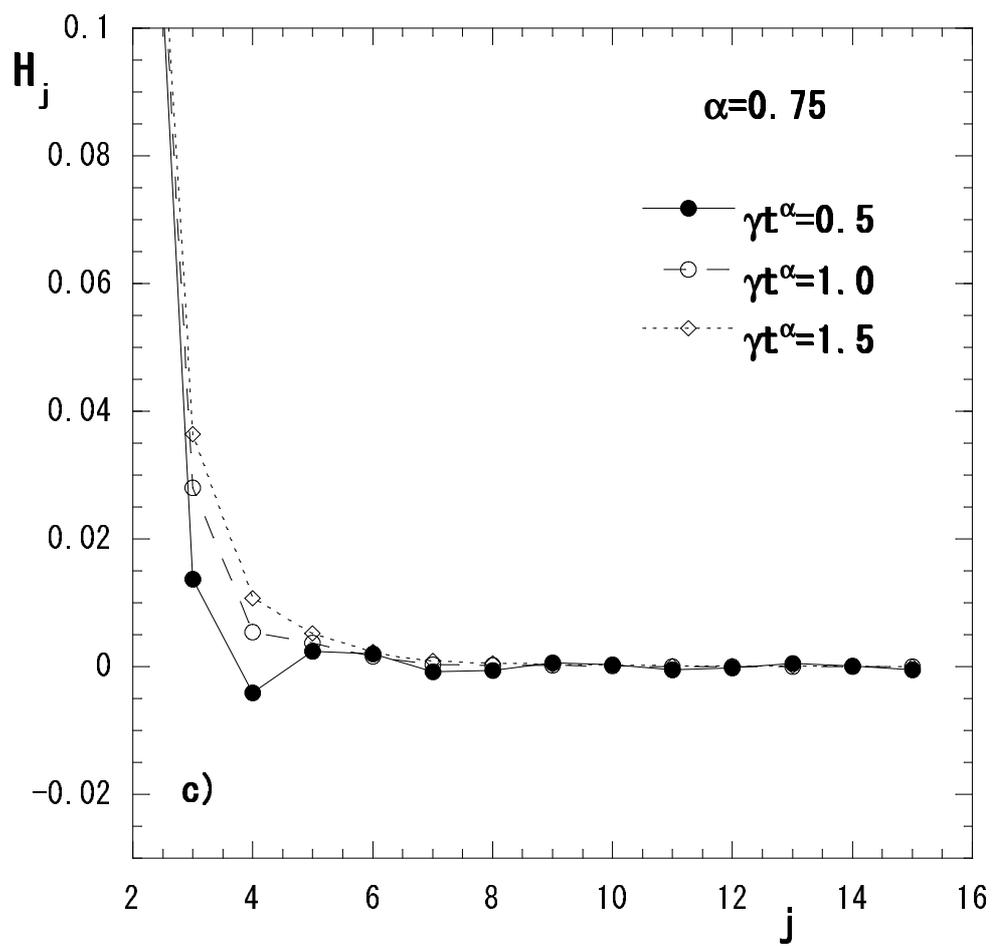
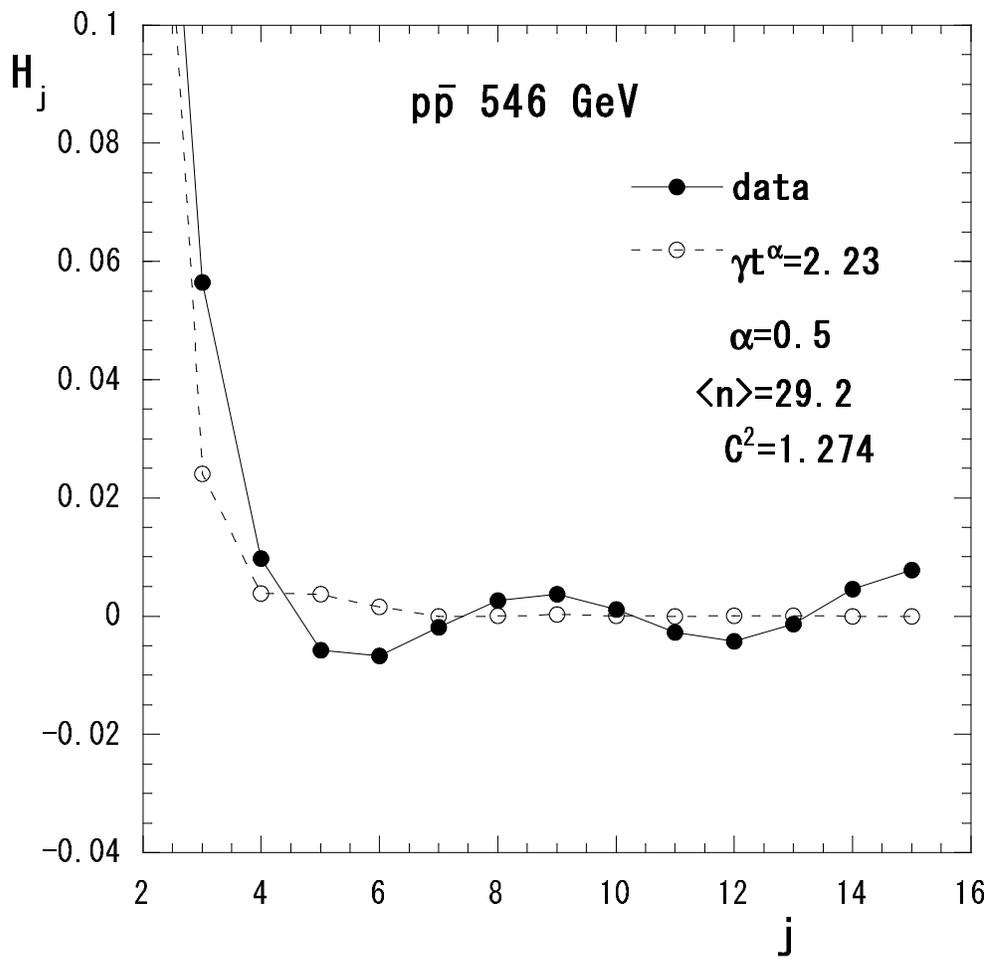
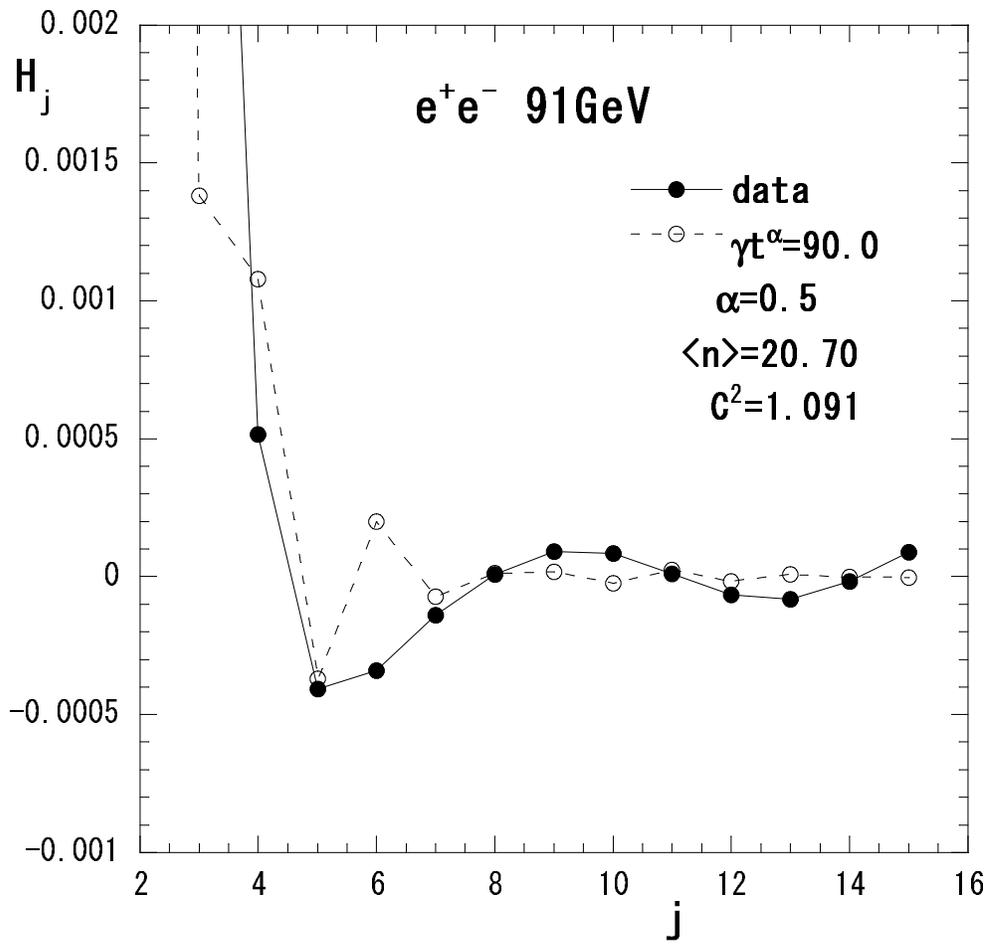
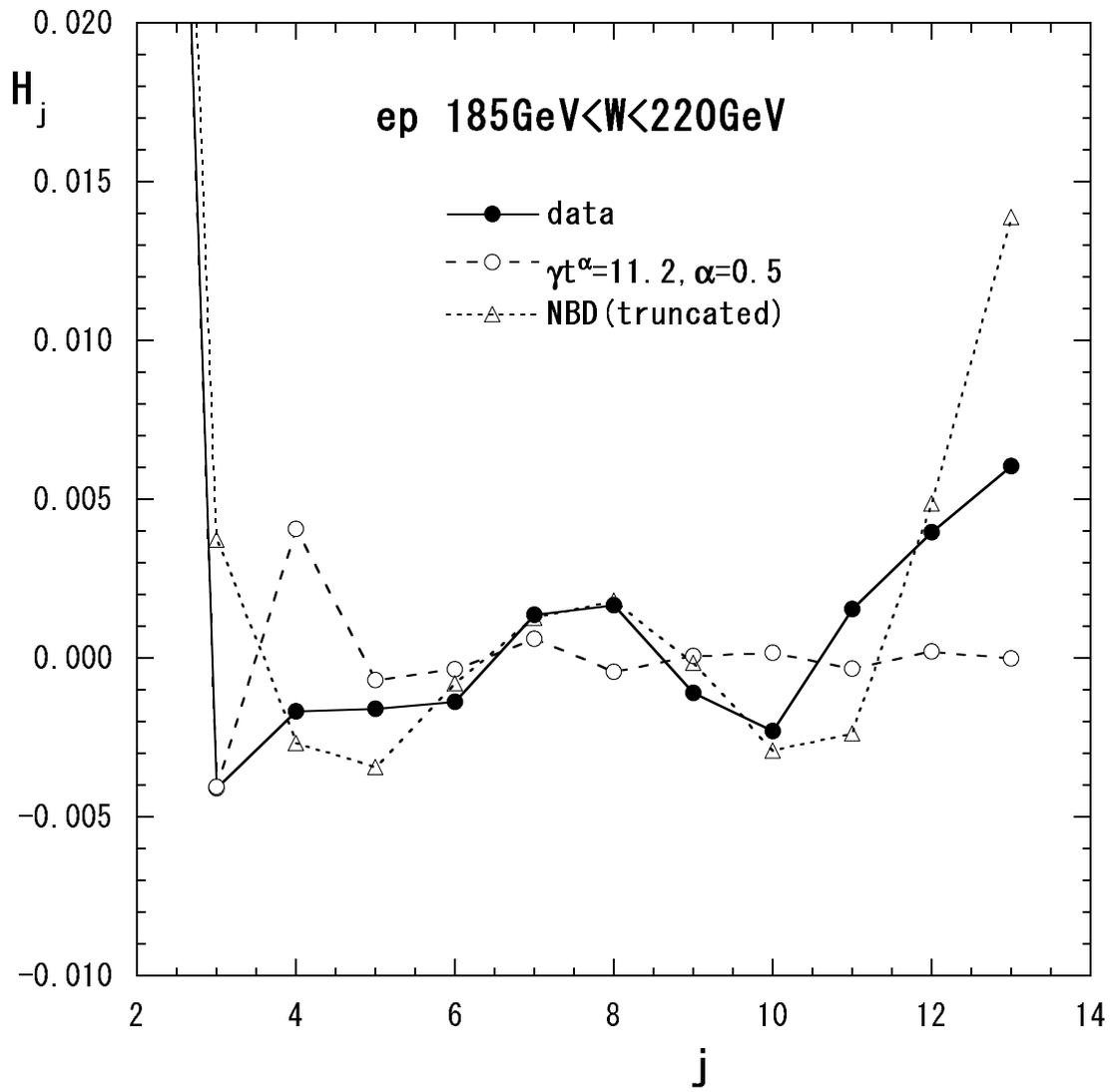
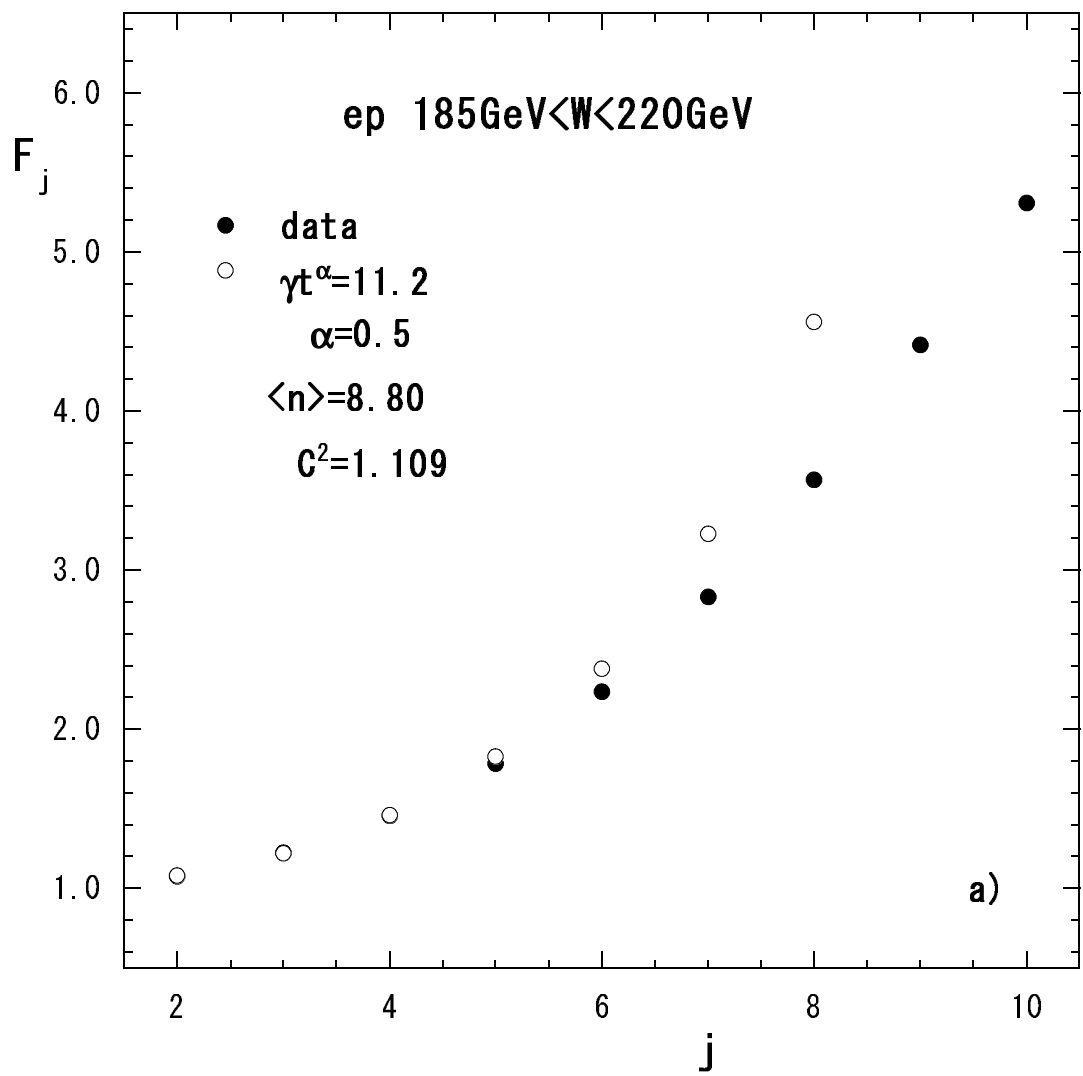
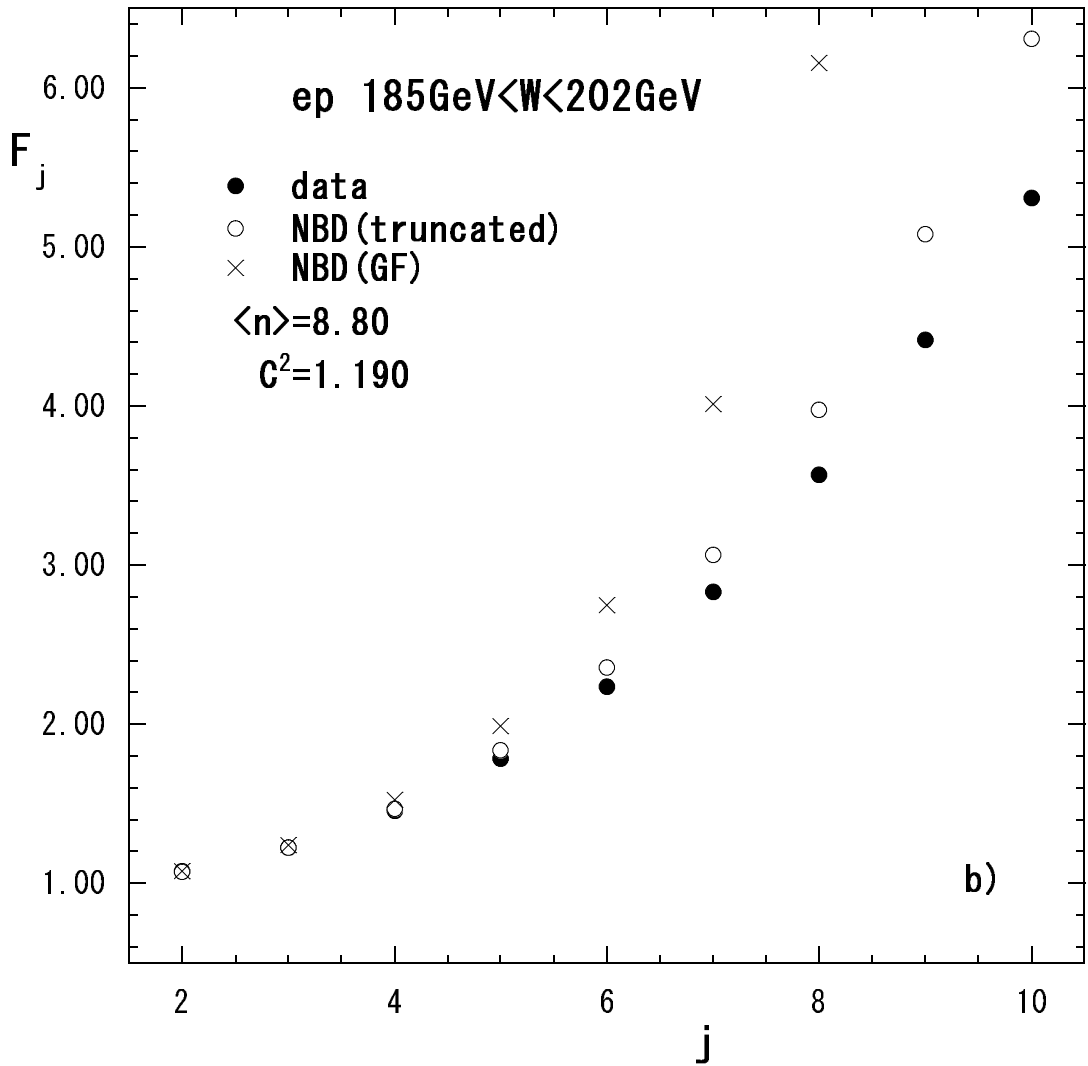
\end{document}